\documentstyle[prl,aps]{revtex}
\font\Bbb =msbm10  scaled \magstephalf
\def\id{{\hbox{\Bbb I}}}

\def\textbf#1{{\bf #1}}
\def\be{\begin{equation}}
\def\ee{\end{equation}}
\def\ben{\begin{eqnarray}}
\def\een{\end{eqnarray}}
\def\eea{\end{array}}
\def\bea{\begin{array}}
\newcommand{\bei}{\begin{itemize}}
\newcommand{\eei}{\end{itemize}}

\begin{document}
\draft
\twocolumn

\title{Measuring quantum entanglement without prior state reconstruction}

\author{Pawe\l{} Horodecki}

\address{Faculty of Applied Physics and Mathematics,
Technical University of Gda\'nsk, 80--952 Gda\'nsk,
Poland}
\maketitle

\begin{abstract}
It is shown that, despite strong nonlinearity,
entanglement of formation of two-qubit state can
be measured without prior state reconstruction.
Collective measurements on small number of copies
are provided that allow to determine quantum concurrence
{\it via} estimation of only {\it four} parameters.
It is also pointed out that another entanglement measure
based on so called negativity can also be measured in similar way.
The result is related to general problem: what kind of
information can be extracted efficiently from unknown quantum state ?
\end{abstract}


Two-qubit state entanglement is well characterised.
It is the only case where analytical formula
for entanglement of formation  $E_{f}$ \cite{huge}
has been provided \cite{W1} with help of so called concurrence
(see \cite{Wootters} for review).
The simple necessary and sufficient conditions
for the presence of entanglement in two qubit state
is known \cite{Peres,sep} involving
so called positive partial transpose (PPT)
map. However the question is how to detect the presence of
entanglement in {\it unknown} state possibly efficiently i. e.
with minimal number of estimated parameters.
As one knowns PPT test is represented by nonphysical operation.
Concurrence and entanglement of formation are highly nonlinear
functions too. Apparently it seems that they
require full prior state reconstruction in general.
The same might be expected to hold for any other entanglement
measures \cite{measure0,measure1} (see also \cite{measure2}).
Indeed direct detection of entanglement of formation function \cite{SH00,Acin}
has succeeded only for pure two-qubit states and relied on very special
property of the function in that case.
Even then, however, the procedure requires estimation
of more than one observable \cite{SH00}.

Quite recently, using formula of the best structural physical
approximations of hermitian
nonphysical maps \cite{nonlinear},
it has been shown \cite{PPTlab} how to detect violation of PPT
separability test (or any positive map separability test)
experimentally without {\it any} prior knowledge about quantum state
(for the analysis of the interferometric scheme see \cite{Estimator}).

In this work we provide a protocol detecting concurrence
of unknown state by estimation of four parameters in collective
measurements of small (not more than eight) number of copies.
We also point out how to estimate
computable entanglement measure \cite{measure1}
based on negativity \cite{negativity,measure1} in
a similar way. For unknown state this gives quadratic gain
in number of parameters if compared to quantum tomography.
The latter protocol is not restricted to
qubits but is valid for any $d \otimes d'$ systems.
We discuss both presented protocols as far as number of
involved copies are concerned. In particular to estimate
two-qubit entanglement of formation one needs
slightly more number of copies than in
state reconstruction scheme. It is however different for ``computable''
measure \cite{measure1} where not only number of parameters
estimated but also number of pairs involved is less than
what state reconstruction requires. 
Subsequently we shall describe the protocols
and conclude with some discussion.

{\it Measuring two-qubit entanglement of formation .-}
Let us start with general formula for entanglement of formation
$E_{f}$ of given bipartite state $\varrho$ \cite{huge}:
$
E_{f}(\varrho)=\mathop{\mbox{min}}\limits_{\{ p_{i}, \psi_{i} \}}
\sum_{i}p_{i}S(Tr_{A}(|\psi_{i}\rangle \langle \psi_{i}|))
$
where $S(\sigma)=-Tr(\sigma log_{2}\sigma)$ is von Neumann
entropy of state $\sigma$ (counted in bits)
and the minimum is taken over all ensembles $\{ p_{i}, \psi_{i} \}$
such that $\sum_{i}p_{i}|\psi_{i}\rangle \langle \psi_{i}|=\varrho$.
For two qubits the entanglement of formation
has been explicitly calculated \cite{Wootters}
and it amounts to:
\begin{equation}
E_f(\varrho)=h(\frac{1+\sqrt{1-C(\varrho)^2}}{2})
\end{equation}
where Shannon binary information is
$h(x)\equiv- xlog_2 (x) - (1-x)log_2 (1-x)$.
The important state function called concurrence \cite{W1}
\begin{equation}
C(\varrho)=max[\sqrt{\lambda_1}-\sqrt{\lambda_2}-\sqrt{\lambda_3}-
\sqrt{\lambda_4},0]
\end{equation}
involves four real monotonically
decreasing numbers $\{ \lambda_{i}\}$ which are
eigenvalues of (nonhermitian) matrix $\varrho \tilde {\varrho}$
with
\begin{equation}
\tilde{\varrho}= \Sigma\varrho^*\Sigma, \ \ \Sigma=\sigma_{y}\otimes \sigma_{y}
\label{matrix}
\end{equation}
where Pauli matrices act on Alice and Bob qubit respectively
(recall that $\varrho$ is two qubit state) and star stands
for complex conjugate.
Obviously neither $C$ nor $E_{f}$ is measurable
in usual quantum mechanical sense i. e. neither of them is an
observable. Indeed this is forbidden
by the very fundamental laws of quantum mechanics:
both functions are nonlinear in state parameters while only {\it linear}
operations can be performed on single copy.
Still one can try to {\it estimate} them in another way -
involving access to several copies.
Such approach succeeded for pure two-qubit states \cite{Acin}
based, however, on the very special property of any pure state
ie. the dependence of its entanglement on eigenvalues
of reduced density matrix.

We would have $C$ and $E_{f}$ determined for
arbitrary given $\varrho$ if only we knew the
four numbers: $\lambda_1$, $\lambda_2$,
$\lambda_3$, $\lambda_4$.
They come from rather complicated nonlinear function
of $\varrho$. Surprisingly for any unknown
$\varrho$ they can be physically measured
by estimating only  {\it four} parameters
(instead of fifteen ones describing full density matrix)
if joint measurements are available.
The idea of the present method is to find
(in collective quantum measurements)
four ``moments'' that allow us to reconstruct $\lambda$-s and,
consequently, concurrence completely.
The basic scheme of the corresponding protocol is very simple:
\begin{eqnarray}
&&\varrho \otimes \varrho
\mathop{\longrightarrow}\limits^{\Lambda_1} \varrho_1
\ \mathop{\Longrightarrow}\limits^{}
 \ \langle M_{1} \rangle =
 \sum_{i}\lambda_{i} \nonumber \\
&&\varrho \otimes \varrho  \otimes
\varrho \otimes \varrho
\mathop{\longrightarrow}\limits^{\Lambda_2} \varrho_2
\ \mathop{\Longrightarrow}\limits^{}
\ \langle M_{2} \rangle = \sum_{i}
(\lambda_{i})^{2}
\nonumber \\
&&\varrho \otimes \varrho  \otimes
\varrho \otimes \varrho  \otimes
\varrho \otimes \varrho
\mathop{\longrightarrow}\limits^{\Lambda_{3}} \varrho_3
\ \mathop{\Longrightarrow}\limits^{}
\langle M_{3}\rangle = \sum_{i}
(\lambda_{i})^{3}
\nonumber \\
&&
\varrho \otimes \varrho  \otimes
\varrho \otimes \varrho  \otimes
\varrho \otimes \varrho  \otimes
\varrho \otimes \varrho
\mathop{\longrightarrow}\limits^{\Lambda_4} \varrho_4
\mathop{\Longrightarrow} \limits^{} \langle M_{4} \rangle
= \nonumber \\
&& \phantom{XXXX} = \sum_{i} (\lambda_{i})^{4}
 \ \ \
\phantom{XX}
\nonumber \\
\label{scheme}
\end{eqnarray}
According to the above, given sample of many copies
of given $\varrho$ we first divide it into four groups
consisting of two, four, six and eight copies
respectively. Then we subject each group to some
specific quantum channel (i. e. completely
positive tracepreserving map) $\Lambda_{k}$, $k=1,2,3,4$.
As a result we get quantum states $\varrho_{k}$.
Finally we estimate mean values
$\langle M_{k} \rangle\equiv Tr(\varrho_{k}M_{k})$
of four observables $M_{k}$.
They represent the ``moments'' needed to reconstruct
values $\{ \lambda_{i} \}$. Moreover, as we shall see below,
each $\langle M_{k} \rangle$ can be determined {\it via} estimation
of single parameter in binary POVM.

To see how the above scheme works we need to recall
the concept of physical approximation of
PPT operation in lab \cite{PPTlab}.
Namely the tracepreserving map
\begin{equation}
\Lambda_{XY} =\frac{d}{d^3 +1} I_{X} \otimes I_{Y} +
\frac{1}{d^3 +1} \id_{X} \otimes T_{Y}, \
\label{map}
\end{equation}
is physically implementable on any bipartite $d \otimes d$
state $\sigma_{XY}$ defined on ${\cal H}_{X} \otimes {\cal H}_{Y}$.
Thus $\Lambda_{XY}$ corresponds to what in literature is
called {\it quantum channel}.
In formula (\ref{map}) $I_{X},I_{Y}$ stand
for identity matrices corresponding to subsystem
$X$, $Y$ respectively. The map $\id$ is just identity
map on subsystem $X$ while $T$ stands
for transposition on second subsystem $Y$.
The partial transposition $\id_{X} \otimes T_{Y}$ simply
transposes indices corresponding to the second subsystem.
The map (\ref{map}) is called {\it structural} physical approximation
of the nonphysical PPT map $\id \otimes T$  \cite{nonlinear,PPTlab}.
In particular one can define in similar way
such approximation for transposition map $T$.
For two qubits the latter coincides with previously introduced
universal-NOT gate (see \cite{UnivNOT}).
Subsequently  detailed description of the scheme (\ref{scheme})
will be presented in two steps.

{\it Step I. Action of channels $\Lambda_{k}$ .-}
We define quantum channel $\Lambda_{k}$ as being
composed of two physical operations:
(i) the map
(\ref{map}) with $d=d_{k}\equiv 4^{k}$
and spaces ${\cal H}_X$ (${\cal H}_Y$) describing
all the copies on odd (even) position in the $k$-th group
and (ii) the subsequent action of unitary operation $\Sigma$
on each ``even'' copy.
For instance in the second group  $\varrho \otimes \varrho
\otimes \varrho \otimes \varrho$ ($k=2$)
the first and the third copy corresponds to subsystem $X$,
while the remaining two represent subsystem $Y$.
Only copies belonging to $Y$ will be subject to subsequent
action (ii) of unitary operation $\Sigma$.

Let the channels $\Lambda_{k}$ act on the 
groups of states according to the scheme (\ref{scheme}).
As a result we get the following four states
\begin{equation}
\varrho_{k}=\frac{d_{k}}{d_{k}^3 +1} I_{X} \otimes I_{Y} +
\frac{1}{d_{k}^3 +1} (\varrho \otimes \tilde{\varrho})^{\otimes k}, \ k=1,...,4.
\label{si}
\end{equation}
Up to the maximal noise part, these states 
can be considered as  ``proportional'' to
the matrices $(\varrho \otimes \tilde{\varrho})^{\otimes k}$.
In fact they differ from the matrices by shrinking factor 
$\frac{1}{d_{k}^3+1}$ at the ``Bloch''.
This property is similar to what happens in the theory
of approximate cloning or in the case of universal NOT gate \cite{UnivNOT}.

{\it Step II. Measurement  of moments $\langle M_{k} \rangle$ .-}
We shall apply the approach \cite{nonlinear,Estimator} providing
spectrum state estimation with help of observables
constructed from shift operators $V_{n}$ \cite{shift}.
They have the following property
\begin{equation}
Tr(V_{(n)} A_{1} \otimes ... \otimes A_{n})=
Tr(A_{1}A_{2}... A_{n})
\label{property}
\end{equation}
Combining the above property with the fact that all moments
$\sum_{i} (\lambda_{i})^{k}$ are real one can see that the latter are
just by mean values of the following observables
\begin{equation}
M_{k}=\frac{d_{k}^3 +1}{2}(V_{(2k)}+ V_{(2k)}^{\dagger})-d_{k}^3 I
\label{mk}
\end{equation}
when each of them is calculated on state $\varrho_{k}$
(see the scheme (\ref{scheme})).
Finally, the numbers $\lambda_{k}$ can be calculated
uniquely from the moments. This can be shown
in analogy to the analysis of Ref.  \cite{nonlinear}
where state spectrum estimation was provided.

Now an important point is that each mean $\langle M_{k} \rangle$
can be detected (up to rescaling) as {\it a single parameter} in
special binary POVM: following Ref. \cite{Estimator} each
observable $M_k$ can be encoded into some ancilla and then
its mean value can be reproduced
form measurement of elementary  binary
observable (Pauli matrix $\sigma_{z}$)
\cite{Koment}.
This concludes the description of the protocol.


The presented method is parametrically
efficient - it requires much less parameters to be measured
than state reconstruction does (four instead of
fifteen). It is not so as long as number of
copies is concerned: one ``round'' of quantum tomography
requires 15 copies
while here (see (\ref{scheme})) we require 2+4+6+8=20 copies
in each round of the protocol.
However the factor $r=r_{p} \cdot r_{c}\equiv$``number of parameters
$\times$ number of copies'' is here superior ($r=80$)
than for the state reconstruction schemes ($r=165$).
Even more striking is the fact that while the number
of copies consumed in one round of the experiment is increased
only by $33 \%$  (from 15 to 20), the number of parameters
is decreased almost four times (from 15 to 4).
It suggests that the optimal method
of entanglement estimation should exists that consumes
the same number of copies as state reconstruction but requires
less number of parameters (5 or 6).
Note that the present method is useful when we are allowed
to use small number of apparata with {\it fixed} architecture.
One can also anticipate the existence of trade-off between
$r_p$ and $r_c$ involved in estimation of any entanglement.
The optimisation of this trade-off is an interesting open problem.

Note that so far we have asked ``How much entanglement is
in the system?''. The above measurement protocol
not only answers but, obviously, solves also the less detailed
problem: ``Is there any entanglement in the system at all ?''.
One can however discern the two questions above
and consider the scenario when the observer
knows that two qubit entanglement is present in the system but
has no idea ``how much''.
Below we shall provide protocol for such a scenario
basing on the fact \cite{WellensKus} that  if $2 \otimes 2$ state $\varrho$
is entangled then the smallest eigenvalue of the matrix
$\gamma\equiv \Sigma \varrho^{T_{A}} \Sigma \varrho^{T_{B}}$
is proportional to $\frac{C(\varrho)^{2}}{4}$ .
The modified protocol must involve the map (\ref{map}) (with
transposed subsystems suitably permuted) to produce the
new states $\gamma_{k}'$ with Bloch vector ``proportional'' to
$(\Sigma \varrho^{T_{A}} \Sigma \otimes \varrho^{T_{B}})^{\otimes k}$
(that play the role of states $\varrho_{k}$ form {\it Step I}).
Then one should estimate eigenvalues of  $\gamma_k$ in the
same way as in Step II and infer $C(\varrho)$
following result of Ref. \cite{WellensKus}.
The above protocol answers quantitative ``how much''
question provided that observers has qualitative
knowledge that entanglement is present.
The latter must be established before and to this
end physical application \cite{PPTlab} of PPT test can be used.
Due to its binary character (``Yes-No'' answer)
it requires {\it less} measurement precision than
the qualitative stage described above.
We can use it to minimise measurement
effort: if PPT test does not reveal
any entanglement we just abandon the second stage.

{\it Multilevel systems and computable entanglement measure .-}
For general $d \otimes d'$ systems (they can be called
multilevel systems) there is no analytical formula
for entanglement of formation.
However at least
one nontrivial entanglement measure
can be calculated analytically.
This is so called ``computable''
entanglement measure \cite{measure1}:
\begin{equation}
E_{c}(\varrho)=log_{2}||\varrho^{T_{B}}||
\label{comput}
\end{equation}
where $\varrho^{T_{B}}=[\id \otimes T](\varrho)$
and $|| \cdot ||$ stands for trace norm which for hermitian
operator means sum of moduli of its eigenvalues.
In particular the quantity $(||\varrho^{T_{B}}||-1)$ introduced to
quantify entanglement \cite{negativity} and after
normalisation $N=(||\varrho^{T_{B}}||-1)/2$
called {\it negativity} has been shown \cite{measure1}
to be {\it entanglement monotone} \cite{Vidal}.
The measure (\ref{comput}) is weaker than entanglement
of formation, it ``detects'' however all
free ie. distillable entanglement of bipartite systems.
To estimate the measure we must experimentally determine
the quantity
\begin{equation}
||\varrho^{T_{B}}||=\sum_{i}|\lambda'_{i}|,
\label{modulus}
\end{equation}
where $\{ \lambda'_{i} \}$ are eigenvalues
of partially transposed $\varrho$ symbolised by
$\varrho^{T_{B}}$. The above formula
is crucial for subsequent analysis.

We now apply scheme of Ref. \cite{PPTlab} making
however full use of its output rather
than focusing on minimal eigenvalue.
For any bipartite $d \otimes d$ state $\varrho$
the scheme consists of application the map (\ref{map})
which leads to the new quantum state
$\varrho_{1}=\Lambda_{XY}(\varrho)$ with eigenvalues $\{ \lambda_{i} \}$.
Each single $ \lambda_{i}$ one eigenvalue
$\lambda'_{i} $ of $\varrho^{T_{B}}$.
Indeed they are related by the following affine function:
\begin{equation}
\lambda_{i}= \frac{d}{d^3 +1} +
\frac{1}{d^3 +1} \lambda'_{i}.
\end{equation}
Now the key point is that  $\{ \lambda_{i} \}$
{\it can be measured} experimentally
{\it via} measurement of $r_{p}=d^{2}-1$
observables (see \cite{nonlinear,Estimator,PPTlab}).
It is {\it much less} than $d^{4}-1$ required for quantum
tomography. From results of the measurements one
can calculate all parameters $\lambda'_{i}=(d^{3}+1)\lambda_{i} -d$
and substitute them to the formula (\ref{comput}) which
provides the needed value of the corresponding measure.

The whole scenario can be immediately generalised
to any $d \otimes d'$ system with $d \neq d'$.
Only the parameters in (\ref{map}) will change
slightly and this can be easily calculated with help of
structural approximation formula \cite{nonlinear}.

It is interesting that in the present scenario one has
also gain in number of copies involved
Indeed we have $r_{c}=2+3+...+ d^2=(d^4 +d^2 -2)/2$
instead of $r_{c}=d^{4} - 1$.

{\it Discussion .-}
In conclusion we have provided the first nontrivial 
protocols  of entanglement estimation that require no prior 
knowledge about the states. 
The first presented protocol allows to detect
two-qubit entanglement of formation
with help of joint measurements on small number of systems.
In one round the method requires $33 \%$ of copies more than state
reconstruction (20 instead of 15). However it needs 
{\it almost four
times less} 
parameters (4 instead of 15) than quantum tomography does.
We have also shown how to estimate experimentally
computable measure of entanglement $E_{c}$
based on partial transpose operation.
Applying the protocol of \cite{PPTlab}
we pointed out that the measure can be determined
with help of estimation of only $d^{2}-1$ parameters
instead of $d^{4}-1$ ones required
for state estimation. The scheme provides also gain
in the number of copies consumed: $r_{c}=(d^{4}+d^{2}-2)/2$
instead of $r_{c}=d^{4}-1$ due to reconstruction schemes.

It would be interesting to find similar
protocol determining some parameters of the best
separable approximation of two-qubit quantum states
(for instance entangled part of Lewenstein-Sanpera
decomposition \cite{LS}).
It involves the question how to find experimentally 
eigenvector corresponding to the least negative eigenvalue 
of partially transposed density matrix $\varrho^{T_{Y}}$.
To get the eigenvector with minimal measurement cost 
would be particularly interesting because it plays crucial role
in the only universal protocol of two-qubit entanglement
distillation \cite{pur}. In the above context basic questions
naturally arise: (i) whether it is possible to estimate eigenvector
corresponding to extreme  eigenvalue
of given density matrix without prior reconstruction
of the latter ? (ii) if so, how efficiently can
it be done ?  Any possible solution of these problems must
take into account fundamental restrictions on
quantum ''nonlinear operations'' \cite{nonlinear,Fiurasek}.

In any case the present results lead naturally to another interesting question:
is it possible to estimate efficiently other entanglement measures
(like relative entropy of entanglement \cite{relative})
or - more generally - correlation measures like the one from
Ref. \cite{correlations}) based of von Neumann entropy ?
For pure states it can be done as all measures
are equivalent to the entropy of the reduced density matrices
which at the same time is proportional to index of correlation
$I=S_{A} + S_{B}- S_{AB}$. For mixed states the latter serves
rather as a measure of global (quantum + classical) correlations
\cite{correlations}. Following the present analysis
the latter can be efficiently measured without state
reconstruction in general. At present one can not develop
the method for the relative entropy of entanglement
or distillable entanglement because
existence of analytical (or partially analytical)
formulas for those measures is still an open issue.

Furthermore from the point of view of general quantum information theory very
important problem is: {\it what kind of information
(whatever it means) can be extracted from 
unknown quantum state at small measurement cost?}.
Here we have considered extraction of information 
represented by special parameter -
entanglement measure. It is closely related to the following 
problem  {\it what kind of measurement information can be extracted from 
the state in the protocols destroying as little quantum 
information as possible ?}
Indeed measurement of small number of parameters
can be always interpreted as an incomplete von Neumann measurement
which destroys  quantum information less than the complete measurement does.
An illustrative example is a detection of  spectrum 
without state reconstruction  via Young diagrams   
projections proposed in Ref. \cite{KeylWerner}
(for alternative spectrum detection  see  \cite{Estimator}). 
The above issues are crucial for some quite fundamental
processes like universal quantum compression.
For example it has been shown \cite{Jozsa} that quantum
source can be compressed efficiently if only
one knowns  {\it single} parameter represented by von Neumann
entropy of the source. The entropy can be
estimated {\it without prior knowledge} about the state:
the measurement effort (see \cite{Estimator}, cf. \cite{KeylWerner})
is the same as in the estimation of the computable
measure considered above.
Finally, we have considered one of the basic problems 
in general entanglement theory: 
''how one can detect the amount of entanglement 
experimentally?''.  This question
concerns not only entanglement itself: any answer to it 
says something about properties of  
information contained in unknown quantum states.

The author thanks A. Ekert, J. Domsta, S. Huelga, M. Plenio
and K. \.Zyczkowski for helpfull discussions. The work is supported
by the IST project EQUIP, contract No. IST-1999-11053.


\begin{references}
\bibitem{shift}
Shift on $({\cal C}^{m} )^{\otimes n}$ is defined as follows: $V_{(n)}e_{1} \otimes e_{2} \otimes ... \otimes e_{n}=
e_{n} \otimes e_{1} \otimes ... \otimes e_{n-1}$
for any basis vectors $\{ e_{l} \}$.
\bibitem{Koment}
Alternatively ''mean''  of  $V_{k}$ can be detected directly \cite{Estimator}.
\end{references}
\end{document}